\documentclass[aps,twocolumn,amsmath,floatfix,byrevtex,prb,nobibnotes]{revtex4}
\usepackage{graphicx}

\begin{document}
\title{The influence of ultra-fast laser pulses on electron transfer in molecular wires studied by a non-Markovian density matrix approach}
\author{Sven Welack} \email{sven.welack@s2000.tu-chemnitz.de}
\author{Michael Schreiber}
\author{Ulrich Kleinekath\"ofer}
\affiliation{Institut f\"ur Physik, Technische Universit\"at Chemnitz, 09107
  Chemnitz, Germany}

\date{\today}
\begin{abstract}
  New features of molecular wires can be observed when they are irradiated
  by laser fields. These effects can be achieved by periodically
  oscillating fields but also by short laser pulses. The theoretical
  foundation used for these investigations is a density matrix formalism
  where the full system is partitioned into a relevant part and a thermal
  fermionic bath.  The derivation of a quantum master equation, either
  based on a time-convolutionless or time-convolution projection-operator
  approach, incorporates the interaction with time-dependent laser fields
  non-perturbatively and is valid at low temperatures for weak system-bath
  coupling.  From the population dynamics the electrical current through
  the molecular wire is determined. This theory including further
  extensions is used for the determination of electron transport through
  molecular wires.  As examples, we show  computations of coherent
  destruction of tunneling in asymmetric periodically driven quantum
  systems, alternating currents and the suppression of the directed current
  by using a short laser pulse.
\end{abstract}


\maketitle

\section{Introduction}
Intra- and intermolecular electron transfer has been studied for several
decades \cite{marc56,poll96,bixo99} including its coherent control
\cite{manc02a}.  Lately the closely related field of molecular electronics
has attracted much interest, especially the transport through molecular
wires\cite{nitz03a,ghos04}. Experimental progress
\cite{reed97,smit02b,reic02} made in this field also spurred a large
theoretical effort \cite{nitz03a,ghos04}.  Many of those theoretical
studies are based on a tight-binding model for the wire.  The
current-voltage characteristics is calculated either using a scattering
approach \cite{muji94a} or an electron transfer scheme \cite{nitz01} while
both formalisms have the same roots \cite{nitz01b}.  The well-known
Landauer-B\"uttiker scattering formalism \cite{land57,butt86} provides a
method to compute the steady-state currents in systems on the nanoscale
connecting two or more electrodes and later extensions of the formalism
also deal with oscillating fields\cite{datt92, win93}.  Many theories are
based on the assumption of a weak coupling of the wire to the leads and
employ a perturbation theory in this small parameter
\cite{kohl04a,li05,cui05,ovch05a}.  In this aspect these approaches are very
similar to Redfield-like approaches in the field of dissipative quantum
dynamics.  Furthermore, in some theories the wire is coupled to a
dissipative environment to mimic relaxation and decoherence in the wire
\cite{guti04,cize04a,cize04b,lehm04a} or to determine current-induced light
emission\cite{galp05}. Additional effects are observed when the molecular
wire is irradiated by a periodic laser field
\cite{lehm03a,lehm03b,lehm04a,kohl04a}.  Only very little is known for
non-periodic driving fields \cite{ovch05a,kurt05}. One of these studies
\cite{kurt05} uses an extension of density functional theory to describe
time-dependent transport phenomena.

In this contribution we are also interested in a description of the current
through a molecular wire which is based on a quantum master equation.  The
aim is to extend those methods and results obtained for periodic laser
fields \cite{lehm03a,lehm03b,lehm04a,kohl04a} to the area of non-periodic
ultra-fast laser pulses. To do so we borrow techniques developed in the
area of dissipative quantum dynamics. Instead of coupling the system, i.e.\ 
the wire, to a bosonic heat bath it is coupled to a fermionic particle
reservoir with which it can exchange particles.  Externally applied optical
fields can influence the dynamics in a direct manner by changing the wire
part but also in an indirect way by influencing the wire-lead coupling. In
the area of dissipative quantum dynamics these processes have been mainly
studied for monochromatic laser fields
\cite{dach94,geva95,goyc96,cuki97,grif98} but also for short laser pulses
\cite{schi98,meie99}. While for monochromatic fields one often employs
Floquet states, as for the transport in periodically driven wires
\cite{lehm03a,lehm03b,lehm04a,kohl04a}, the situation is more complex for
non-periodic laser pulses.  Meier and Tannor \cite{meie99} proposed a
method how to use a special parametrization of the so-called spectral
density of the reservoir leading to a set of coupled equations for a
primary and several auxiliary density matrices. Within this approach one
automatically accounts for the influence of the external field on the
dissipation operator.  The technique of Meier and Tannor is based on the
time-nonlocal (TNL) Nakajima-Zwanzig identity \cite{naka58,zwan61}, while a
similar scheme with auxiliary operators has been developed also for
time-local (TL) quantum master equations \cite{xu02,klei04a}.  A similar
technique, i.e.\ a direct decomposition of the correlation function, was
introduecd by Korolkov and Paramanov in earlier work \cite{koro97}.

In the current work the technique of auxiliary density matrices is used in
the framework of molecular wires. The model Hamiltonian employed for this
purpose is introduced in the following section.  Section III shows the
evolution of the density matrix using a decomposition of the spectral
density while Section IV is dedicated to the derivation of the current
equation used for the numerical calculations. In Section V, we will make a
few comments on TNL versus TL theory. The formalism is tested for periodic
and non-periodic laser fields in section VI. The last section gives a
summary and outlook.

\section{Model}

\begin{figure}
\includegraphics[width=6cm,clip]{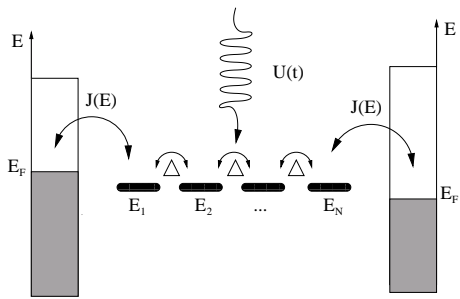}
\caption{A multi-site system is studied in which the two outermost sites
  are attached
  to electronic reservoirs in thermal equilibrium with their respective
  Fermi energies $E_{F,l}$ and $E_{F,r}$. The coupling of the outermost
  sites to the corresponding lead is described by a spectral density
  function $J_R(\omega$) and the different sites are connected to each
  other by a hopping element $\Delta$. The on-site energies $E_i$ of the
  wire can be manipulated with a time-dependent external electric field
  $U(t)$.
\label{fig:1}}
\end{figure}

The system of interest can be represented in a very general form by a
time-dependent Hamiltonian
\begin{equation} \label{equ:Ham_total}
H(t)=H_S(t)+H_R+H_{SR},
\end{equation}
which consists of the time-dependent part describing the relevant system
$H_S(t)$, a Hamiltonian describing the environment $H_R$, in the presented
study the electronic leads, and a coupling term $H_{SR}$ between the
relevant system and its environment.  The wire is formed by electronic
sites $n$ coupled to each other by a hopping element $\Delta$. In second
quantization, this orbital tight-binding description of the molecular wire
reads
\begin{equation} \label{equ:Ham_wire}
H_S(t)=\sum_{nn'} H_{nn'}(t) c_n^\dagger c_{n'}
\end{equation}
where $c_n$ annihilates and $c^\dagger_n$ creates an electron at site $n$
with the anticommutator $[c_n^\dagger,c_{n'}]_+=\delta_{n,n'}$.  Here, the
time-dependence of $H_S(t)$ occurs only in to the irradiation of the system
by an external electromagnetic field that manipulates the on-site energies
$E_n$ with a time-dependent on-site potential $U_n(t)$.  By neglecting
possible influences of the external field on the tight-binding hopping
element $\Delta$, the tight-binding matrix elements $H_{nn'}(t)$ can be
decomposed into
\begin{equation}\label{equ:Hamiltonian_dec}
H_{nn'}(t) = -\Delta (\delta_{n+1,n'}+\delta_{n,n'+1}) +(E_n+U_n(t))\delta_{nn'}.
\end{equation}
The environment of the wire consists of two electronic leads that are
modeled by two independent electron reservoirs in thermal equilibrium. For each
lead, the Hamiltonian $H_R$ in second quantization is given by
\begin{equation} \label{equ:Ham_lead}
H_R=\sum_q \omega_q c_q^\dagger c_q
\end{equation}
with $c_q^\dagger$ and $c_q$ creating and annihilating an electron in the
corresponding reservoir mode $\vert q \rangle$ with mode energy $\omega_q$. Due
to the assumed thermal equilibrium of the electronic leads, the occupation
expectation values of the reservoir modes are determined by
\begin{equation} \label{equ:equi}
\langle c_q^\dagger c_{q'} \rangle = n_F(\omega_q-E_F) \delta_{qq'},
\end{equation}
where $n_F$ is the Fermi function and $E_F$ the Fermi energy. The left
electronic lead is coupled to the first site and the right one to the last
site $N$ of the wire. To keep the notation simple, we will only refer to
the left lead in further derivations but the formalism has to be applied
to the right lead as well.  The coupling of the left electronic lead with
the first site of the wire reads in second quantization
\begin{equation} \label{equ:Ham_coup}
H_{SR}= \sum_q (V_q c_1^\dagger c_q + V_q^* c_q^\dagger c_1)
\end{equation}
with a system-lead coupling strength $V_q$ for each reservoir mode. In general,
these coupling values are determined by the electronic bands of the leads,
e.g. the gold contacts, that couple with the energy levels of the wire and
by the occupation level of these bands with electrons given by the Fermi
energy and the corresponding Fermi function.

\section{Reduced density matrix and time-nonlocal approach}
\subsection{Equation of motion}
In general, the equation of motion (EOM) for the complete density operator $\rho(t)$ including the wire and the leads is given by the Liouville equation
\begin{equation}
\frac{\partial \rho(t)}{\partial t} = -\frac{i}{\hbar} [H(t), \rho(t)] = -\frac{i}{\hbar} \mathcal L(t) \rho(t).
\end{equation}
But the information of interest is limited only to the system part of the
density operator $\rho_S(t)$, i.e. the wire part, which can be obtained by
defining a projection operator $P$ projecting the complete system onto the
relevant part and by tracing out the reservoir degrees of freedom $P
\rho(t)=\rho_R \, \mathrm{tr}_R\lbrace\rho(t) \rbrace = \rho_R \otimes
\rho_S(t)$. The irrelevant part can be determined by the projection
$Q=1-P$. Solving the EOM for the irrelevant part to express the EOM of the
relevant part leads directly to the Nakajima-Zwanzig
\cite{naka58,zwan61,zwan64} operator identity
\begin{eqnarray}\label{equ:NZ}
P\dot{\rho}&=&-i P \mathcal L(t) P \rho(t)\\
& &-P \mathcal L(t) \int_{t_0}^t \mathrm d t' \,\vec T e^{-i \int_{t'}^{t} \mathrm d \tau Q \mathcal L(\tau)}
Q \mathcal L(t') P \rho(t'), \nonumber
\end{eqnarray}
where the initial-value term is neglected. Since all the operators are
chronologically ordered in time, the literature refers to this TNL approach
often as chronological time ordering prescription
(COP)\cite{muka78,reic97,yan98} or time convolution approach \cite{breu99}.
In order to derive an applicable method to solve Eq.\ (\ref{equ:NZ}), we
apply second order perturbation theory \cite{haak73,blum96,meie99} and
trace over the reservoir degrees of freedom to get
\begin{eqnarray} \label{equ:NZtraced}
\dot{\rho}_S(t)&=& -i \mathcal L_S(t) \rho_S(t) \nonumber \\
& & -\mathrm{tr}_R \left\{ \mathcal L_{SR}(t) \int_{t_0}^t \mathrm dt' 
U_{0}(t,t') \mathcal L_{SR}(t') \rho(t') \right\}. \,\,\,\,
\end{eqnarray}
Here $\mathcal L_S$, $\mathcal L_R$ and $\mathcal L_{SR}$ are the Liouville
operators which, respectively, apply $H_S$, $H_R$ and $H_{SR}$.
$U_{0}(t,t')=\vec{T}e^{-\frac{i}{\hbar} \int_{t'}^t \mathrm d\tau
  \left[\mathcal L_S(\tau)+\mathcal L_R\right]} $ is the time evolution
operator of the full system without the interaction part $H_{SR}$ and
$\vec{T}$ is the time ordering operator in positive time direction.  The
annihilation and creation operators of the system and the environment are
defined in different Hilbert spaces which makes it possible to rewrite the
coupling Hamiltonian in Eq.\ (\ref{equ:Ham_coup}) as
\begin{equation} \label{equ:coup_separation}
H_{SR}= \sum_{x=1}^2 K_x \otimes \Phi_x
\end{equation}
with $\Phi_1=\sum_q V_q c_q$, $\Phi_2=\sum_q V_q^* c_q^\dagger$ and $K_1=c_1^\dagger$, $K_2=c_1$.
Inserting Eq.\ (\ref{equ:coup_separation}) in Eq.\ (\ref{equ:NZtraced}) and using the definition of the Liouville operators, a non-Markovian quantum master equation in the Schr\"odinger picture is obtained \cite{may00,li05,cui05,klei04a} 
\begin{eqnarray}\label{equ:master1}
\dot{\rho}_S(t)&=& -i \mathcal L_S(t) \rho_S(t) \nonumber \\
& &-\int_{t_0}^t \mathrm{d}t' \sum_{xx'} \mathrm{tr}_R \left\{ 
[K_x \Phi_x, U_s(t,t') e^{-i H_R (t-t')} \right. \nonumber \\
& & \left. \times{}[K_{x'} \Phi_{x'}, \rho_S(t',t_0)] e^{i H_R(t-t')}]  \right\},
\end{eqnarray}
with $U_S(t,t')=T_+e^{-\frac{i}{\hbar} \int_{t'}^t \mathrm d\tau \mathcal L_S(\tau)} $ being the time evolution operator of the relevant system.

The trace over the lead degrees of freedom and all the reservoir operators in the dissipation term can be summarized into  correlation functions
\begin{equation} \label{equ:bcxx}
C_{x x'}(t)=\mathrm{tr}_R \lbrace e^{i H_R t} \Phi_x e^{-i H_R t} \Phi_{x'} \rho_R \rbrace
\end{equation}
that contain all the information about the reservoir and its interaction
with the corresponding wire site. For the system of interest, these
correlation functions decay in time thereby causing a memory loss in the
dissipation term of Eq.\ (\ref{equ:master1}).

Due to the thermal equilibrium condition of the electronic leads, two of
the four functions are zero, $C_{11}=C_{22}=0$. This reduces the summation
over $x$ and $x'$ to the pairs $(xx')=(12)$ and $(xx')=(21)$.  Using the
property of the correlation functions that $C_{x x'}(t)=C_{x
  x'}^*(-t)$, we can define the following auxiliary operators
\begin{equation}\label{equ:aux1}
\Lambda_{xx'}^{}(t) = \int_{t_0}^t \mathrm dt' C_{xx'}(t-t')  U_S(t,t')  K_{x'} \rho_S(t'),
\end{equation}
\begin{equation}\label{equ:aux2}
\widehat \Lambda_{xx'}(t) = \int_{t_0}^t \mathrm dt' C_{x'x}^*(t-t')  U_S(t,t') \rho_S(t') K_{x'}
\end{equation}
to simplify Eq.\ (\ref{equ:master1}). These auxiliary operators incorporate
the memory of the system and weight the time-dependent electron transfer
between the wire and the lead. Expressing the quantum master equation
(\ref{equ:master1}) in terms of the auxiliary operators, we get the final
expression for the master equation of the reduced density operator at time
$t$
\begin{equation}\label{equ:master2}
\dot{\rho}_S(t)=-i \mathcal L_S(t) \rho_S(t) - \sum_{xx'} [K_x,\, \Lambda_{xx'}(t)-\widehat \Lambda_{xx'} (t)].
\end{equation}

\subsection{Spectral decomposition}
In analogy to methods recently developed for systems coupled to a bosonic
bath\cite{meie99,li05,cui05,klei04a}, we develop equations of motions for the
auxiliary operators defined by Eqs.\ (\ref{equ:aux1}, \ref{equ:aux2}) using a
numerical decomposition of the spectral density $J_R(\omega)$ to decompose
the reservoir correlation functions. Solving the trace in Eq.\ (\ref{equ:bcxx})
and making use of Eq.\ (\ref{equ:equi}), the nonvanishing correlation
functions read
\begin{equation}
C_{12}(t)= \sum_q \vert V_q \vert^2 n_F(-\omega_q +E_F) e^{-i \omega_q t}
\end{equation}
\begin{equation}
C_{21}(t)=\sum_q \vert V_q \vert^2 n_F(\omega_q -E_F) e^{i \omega_q t}.
\end{equation}
All the external properties of the fermionic reservoir are described by a single quantity, namely the spectral density $J_R(\omega)$, which can be generated by a superposition of weighted $\delta$ functions
\begin{equation} \label{equ:spectralgeneral}
J_{R}(\omega)=\sum_q \pi \vert V_q \vert^2 \delta(\omega-\omega_q).
\end{equation}
Equation (\ref{equ:spectralgeneral}) becomes a smooth function for a dense
spectrum of the reservoir modes.  Using the properties of the $\delta$ function
to transform the summation over $q$ into an integral over $\omega$ and
extending its lower limit to $-\infty$ by assuming $J_{R}(\omega)=0$ for
$\omega < 0$, since we need an improper form of the integrals to solve them
by applying the theorem of residues, we get the final integral equations
for the reservoir correlation functions
\begin{equation} \label{bath12}
C_{12}(t)=\int_{-\infty}^\infty \frac{\mathrm d\omega}{\pi} \, J_{R}(w) \,n_F(-\omega_q +E_F)  e^{-i \omega t}.
\end{equation}
\begin{equation} \label{bath21}
C_{21}(t)=\int_{-\infty}^\infty \frac{\mathrm d\omega}{\pi} \, J_{R}(w) \,n_F(\omega_q -E_F) e^{i \omega t}
\end{equation}
To solve these integrals, we will pursue a method proposed by Meier and Tannor \cite{meie99} for bosonic systems and extend it to fermionic systems using a numerical decomposition of the spectral density for $\omega \geq 0 $
\begin{equation} \label{equ:spectralnum}
J_{R}(\omega)=\sum_{k=1}^m \frac{p_k}{4 \Omega_k}  \frac{1}{(\omega -\Omega_k)^2+\Gamma_k^2},
\end{equation}
with real fitting parameters $p_k$, $\Omega_k$ and $\Gamma_k$. This
decomposition is not restricted to a certain shape of the spectral density
and can therefore be used to approximate complicated band structures. This
enables one to avoid the assumption of the wide-band limit and to take
influences of the band structure on the dissipative electron transfer
between the wire and the leads fully into account.

With the complex roots of the Fermi function and of function (\ref{equ:spectralnum}), the theorem of residues applied to Eqs.\ (\ref{bath12}, \ref{bath21}) results in
\begin{eqnarray} \label{bath12dec}
C_{12}(t)&=&\sum_{k=1}^m \frac{p_k}{4 \Omega_k \Gamma_k}
\left(n_F(-\Omega_k^- +E_F) e^{-i\Omega_k^- t} \right) \nonumber \\
& &-\frac{2i}{\beta} \sum_k^{m'} J_{R}(\nu_k) e^{-i \nu_k t}
\end{eqnarray}
\begin{eqnarray} \label{bath21dec}
C_{21}(t)&=&\sum_{k=1}^m \frac{p_k}{4 \Omega_k \Gamma_k}
\left(n_F(\Omega_k^+-E_F)  e^{i\Omega_k^+ t} \right) \nonumber \\
& &-\frac{2i}{\beta} \sum_k^{m'} J_{R}(\nu_k) e^{i \nu_k t}
\end{eqnarray}
with the abbreviations $\Omega_k^+=\Omega_k+i \Gamma_k$ and $\Omega_k^-=\Omega_k-i \Gamma_k$
and the Matsubara frequencies $\nu_k$ given by $\nu_k=i\frac{2\pi k + \pi}{\beta} +E_F$.
These equations determine the coefficients necessary to finally write the correlation functions as a superpostion of weighted exponential functions: 
\begin{equation}
C_{12}(t)=\sum_{k=1}^{m+m'} a_{12}^k e^{\gamma_{12}^k t}
\end{equation}
\begin{equation}
C_{21}(t)=\sum_{k=1}^{m+m'} a_{21}^k e^{\gamma_{21}^k t}.
\end{equation}
Rigorously, the sum over the Matsubara frequencies would be infinite but it
can be truncated at a finite value depending on the temperature of the
system $T$ and the spectral width of $J_R(\omega)$.
This representation for the correlation function allows us to derive a set
of differential equations for the auxiliary density operators
(\ref{equ:aux1}, \ref{equ:aux2}), viz.

\begin{eqnarray}\label{equ:diffaux1}
\frac{\partial}{\partial t} \Lambda_{xx'}^k(t)&=& a_{xx'}^k  K_{x'} \rho_S(t) -i [H_S(t), \Lambda_{xx'}^k(t)] \nonumber \\
& &+ \gamma_{xx'}^k  \Lambda_{xx'}^k(t),
\end{eqnarray}
\begin{eqnarray}\label{equ:diffaux2}
\frac{\partial}{\partial t}{\widehat\Lambda}_{xx'}^k(t)&=&\left(a_{x'x}^k\right)^\ast \rho_S(t) K_{x'}   -i [H_S(t), \widehat \Lambda_{xx'}^k(t)] \nonumber \\
& &+ \left(\gamma_{x'x}^k \right)^\ast \widehat \Lambda_{xx'}^k(t)
\end{eqnarray}
with ${\Lambda}_{xx'}(t)=\sum_{k=1}^{m+m'} {\Lambda}_{xx'}^k(t)$ and ${\widehat\Lambda}_{xx'}(t)=\sum_{k=1}^{m+m'} {\widehat\Lambda}_{xx'}^k(t)$.
These equations can be solved numerically using a simple Runge-Kutta method without the need of diagonalizing the Hamiltonian and together with Eq.\ (\ref{equ:master2}), one now has a complete set of differential equations describing the population dynamics in the wire coupled to the lead in second-order perturbation theory for an arbitrary time-dependent wire Hamiltonian. Regarding the right lead, one just needs to add a second dissipation term to the master equation (\ref{equ:master2}) with differently defined $K_x$ operators, i.e. acting on the last wire site $N$, and a second corresponding set of differential equations for the auxiliary operators.

\section{Current equation}
An intuitive approach to the electric current equation is to consider the
rate of change of the number of electrons, with elementary charge $e$,
inside the lead which reads in the density matrix formalism
\cite{brud93,lehm02,li05,cui05,ovch05a,lehm02a,lehm03a,lehm03b,lehm04a,cama04,kohl04a,kohl04b}
\begin{equation}\label{equ:ansatz}
I_l(t)=e\frac{\mathrm d}{\mathrm dt} \mathrm{tr} \, \lbrace N_l \rho(t) \rbrace =-ie \, \mathrm{tr} \left\{ [N_l,H(t)] \rho(t) \right\}.
\end{equation}
Here $N_l=\sum_q c_q^{\dagger} c_q$ denotes the electron number operator of
the left lead with the summation performed over the reservoir degrees of
freedom. Similar to the last sections, all calculations refer only to the
left lead but are also valid for the right lead by adding the corresponding
terms to the final set of differential equations. The trace and the density
operator $\rho$ in Eq.\ (\ref{equ:ansatz}) are defined in the Hilbert space
of the full system consisting of wire and lead.  Solving the commutator in
Eq.\ (\ref{equ:ansatz}) and making use of the equilibrium condition of the
fermionic reservoir results in $ I_l(t)=-2e \, \sum_q \mathrm{Im} \,
\mathrm{tr} \lbrace V_q c_1^\dagger c_q \rho(t) \rbrace.  $ Since the trace
is invariant under a transformation into the interaction picture, $\tilde
A(t)= U_S(t,t_0) A$ for any operator $A$ defined in the Hilbert space of
the full system, one can use the integrated form of the Liouville equation
in the interaction picture $ \tilde \rho(t)=\tilde \rho(t_0)- i
\int_{t_0}^t \mathrm dt' \, [\tilde H_{SR}(t'), \tilde \rho(t')] $ for the
time evolution of the full density operator $\tilde \rho(t)$ in the
interaction representation.  After solving all the commutator relations,
separating the system part from the reservoir part and employing again the
definition of the correlation functions (\ref{equ:bcxx}), the ansatz for
the current (\ref{equ:ansatz}) finally reads
\begin{eqnarray}\label{equ:current2}
I_l(t)=2e \, \mathrm{Re}  \int_{t_0}^t  \mathrm dt'  \left( \, \mathrm{tr_S}
\left\{ \tilde c_1^\dagger(t) \tilde c_1(t') \tilde \rho_S(t') \right\} \,C_{12}(t-t') \right. \nonumber \\
\left. -\mathrm{tr_S} \left\{ \tilde c_1 (t') \tilde c_1^\dagger (t) \tilde \rho_S(t') \right\} \, C_{21}^*(t-t') \right). \, \, \, \, \,
\end{eqnarray}
The current information is partly contained in the temporal phase relation between the annihilation and creation operators which can be incorporated into a single time evolution operator $U_S$, defined in the Hilbert space of the relavant system, by using the cyclic properties of the trace to write Eq.\ (\ref{equ:current2}) as
\begin{eqnarray}
I_l(t)=2e \, \mathrm{Re}\, \left( \mathrm{tr_S}
\left\{ c_1^\dagger \int_{t_0}^t  \mathrm dt' \, U_S^\dagger(t',t) \, c_1 \rho_S(t') \right\} \,C_{12}(t-t') \right.  \nonumber \\
\left. -\mathrm{tr_S} \left\{ c_1^\dagger \int_{t_0}^t  \mathrm dt' \,
    U_S^\dagger(t',t) \, \rho_S(t') c_1 \right\} \, C_{21}^*(t-t') \right). \, \, \, \, \,
\end{eqnarray}
These integrals have the same structure as the auxiliary density matrices defined in the last section and can be expressed in terms of $\Lambda_{12}(t)$ and $\widehat\Lambda_{12}(t)$. Thus, we get the final equation for the time-dependent current
\begin{equation} \label{equ:finalcurrent}
I_l(t)=2e \, \mathrm{Re} \left(\mathrm{tr}_S \left\{ c_1^\dagger \Lambda_{12}(t) -c_1^\dagger \widehat\Lambda_{12}(t) \right\} \right)
\end{equation}
between the left lead and the first site of the wire. Similar to the master
equation (\ref{equ:master2}) for the population dynamics, all the
information about the interaction of the system with the reservoir is
contained in the time-dependent auxiliary matrices (\ref{equ:aux1},
\ref{equ:aux2}) weighting the corresponding system operator. Eq.\
(\ref{equ:finalcurrent}) is valid within the framework of assumptions we
have made to derive Eq.\ (\ref{equ:master2}) and allows one to measure
the time-resolved current for arbitrary time-dependent systems.

The EOM for the required auxiliary operators are already known from the
time propagation of the population. Thus, we have a complete set
of differential equations consisting of Eqs.\ (\ref{equ:diffaux1},
\ref{equ:diffaux2}), the master equation (\ref{equ:master2}) and the
current equation (\ref{equ:finalcurrent}) and therefore an applicable
formalism to compute the current and the population dynamics for a wide
range of systems. The same set of equations has to be applied to the last
site of the wire as well to describe a wire enclosed by two leads, as shown
in Fig.\ \ref{fig:1}. Furthermore, the ability to define any desired number
of leads (with respect to the given computational resources) makes it
possible to realize driven current switches with an arbitrarily complicated
configuration.

\section{Alternative time-local approach}
Since we started with the Nakajima-Zwanzig identity to derive our equations
for the population dynamics the final expressions are in a TNL form. In
this section, we shortly present their TL counterparts without giving an
extended comparison of both approaches.  In the literature, the TL approach
is also known as time-convolutionless formalism\cite{breu99}, partial time
ordering prescription\cite{muka78,reic97,yan98} or Tokuyama-Mori
approach\cite{cape94b}.  Using the Tokuyama-Mori identity one gets in
second-order perturbation theory for the wire-lead coupling the following
equations for the population dynamics of the wire,
\begin{eqnarray}\label{equ:master2local}
\dot{\rho}_S(t)&=&-i \mathcal L_S(t) \rho_S(t) \\ \nonumber
& & -\sum_{xx'} [K_x,\, \Lambda_{xx'}(t)\rho_S(t)-\rho_S(t)\widehat \Lambda_{xx'} (t)]
\end{eqnarray}
with the modified corresponding auxiliary operators
\begin{equation}\label{equ:aux1local}
\Lambda_{xx'}^{}(t) = \int_{t_0}^t \mathrm dt' C_{xx'}(t-t')  U_S(t,t')  K_{x'}
\end{equation}
\begin{equation}\label{equ:aux2local}
\widehat \Lambda_{xx'}(t) = \int_{t_0}^t \mathrm dt' C_{x'x}^*(t-t')  U_S(t,t')  K_{x'}.
\end{equation}
One can also derive these equations by just applying the substitution $\rho(t')=U_S^\dagger(t,t')\rho(t)$ to the
corresponding set of TNL equations\cite{klei04a}, neglecting the influence of dissipation during the time propagation
of the density operator within the integral. Thus, the current equation reads\cite{li05,cui05}
\begin{equation} \label{equ:finalcurrentlocal}
I_l(t)=2e \, \mathrm{Re} \left(\mathrm{tr}_S \left\{ c_1^\dagger \Lambda_{12}(t) \rho_S(t)- c_1^\dagger \rho_S(t) \widehat \Lambda_{12}(t)\right\} \right).
\end{equation}
TNL and TL approaches for similar systems have been compared
in recent numerical studies\cite{klei04a,yan05} and it depends strongly on the coupling parameters which of both is favorable. Within the parameter regime we used for the following numerical investigations, TL and TNL calculations showed almost identical results.

\section{Numerical results for the laser driven molecular wire}

The system depicted in Fig.\ \ref{fig:1} represents a simple configuration
which allows one to study a variety of transport phenomena. The following
numerical results are expressed in terms of the tight-binding hopping
parameter $\Delta$, where $\Delta=0.1 \mathrm{eV}$ is a reasonable value
for molecular
systems\cite{lehm02,lehm02a,lehm03a,lehm03b,lehm04a,cama04,kohl04a,kohl04b}.
The system temperature is set to $T=0.25 \Delta / k_B=290 \mathrm{K}$.
Despite the fact that it would be possible in Eq.\ (\ref{equ:spectralnum}),
we do not simulate a realistic spectrum for the coupling between the leads
and the wire, but rather restrict the spectrum to a single Lorentzian. For
example, a realistic gold s-band would be a superposition of different
Lorentzians forming certain band edges. Taking only one Lorentzian into
account corresponds to a rough approximation of only one band edge, but
this makes the system and the underlying physical processes easier to
understand. Due to the weak coupling requirement, the spectral density
should be about one order smaller than the internal dynamics, thus the peak
of the Lorentzian $J_R(\omega)$ is set to $0.1 \Delta$, what is guaranteed
by the condition $p_1=0.1 \left(4\Delta \Omega_1 \Gamma_1^2 \right)$.
Reasonable values for the bandwith parameters $\Gamma_k$ are in the region
of $10\mathrm{eV}$.  With the chosen energy settings, a time unit in the
system corresponds to $0.66 \mathrm{fs}$ what enables one to resolve
time-dependent effects on a femtosecond scale. The resulting current unit
can be extracted from Eq.\  (\ref{equ:finalcurrent}) and is equal to a
macroscopic value of $1 [e]=2.43 *10^{-4} \mathrm{A}$.

As a basis set for the tight-binding Hamiltonian we choose the many-body Fock space in which each site is represented by a two-state vector
$\vert \chi_n \rangle$. The set of these state vectors forms the total state vector $\vert \Psi \rangle$ by the tensor product $
\vert \Psi \rangle =\vert \chi_1 \rangle \otimes \vert \chi_2 \rangle \otimes
 ... \otimes \vert \chi_N \rangle$.
The corresponding annihilation and creation operators for site $n$
\begin{equation}
c^\dagger_n=\left(\begin{array}{cc} 0 & 0 \\ 1 & 0 \end{array} \right) \, , \,\,\,\,\,
c_n=\left(\begin{array}{cc}0 & 1 \\ 0 & 0 \end{array} \right)
\end{equation}
represent a complete operator basis for the master and the current equation.

\subsection{The transient oscillation of the undriven wire}

\begin{figure}
\includegraphics[angle=-90,width=8.5cm,clip]{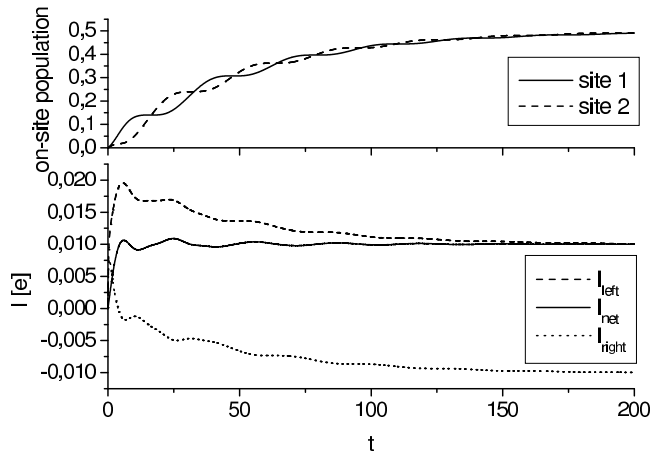}
\caption{Upper panel:   population dynamics measured by $ \mathrm{tr}(c^\dagger_n c_n \rho_S(t))
  $ $(n=1,2)$ for the system with two sites without driving
  field $U(t)$ as a function of time $t$ starting with unoccupied wire
  states. The applied DC voltage is realized by a difference between the
  left and the right Fermi energy of $E_{F,l}-E_{F,r}= 10 \Delta$. The
  on-site energies of the wire are aligned and centered between the left
  and the right Fermi energy $E_1=E_2=E_{F,r}+5\Delta=E_{F,l}-5\Delta$.
  Lower panel: the corresponding currents flowing from the left lead into
  the first site $I_{left}$, from the right lead into right site
  $I_{right}$ and the net current given by
  $I_{net}=(I_{left}-I_{right})/2$.
\label{fig:2}}
\end{figure}

The equilibrium condition of the on-site population of the wire in the
absence of a second lead is determined by the Fermi function $n_F$ taken at
the on-site energy of the site, e.g. $\mathrm{tr}(c_1^\dagger c_1
\rho_S(t))=n_F(E_1-E_F)$. In this particular case, the current drops to
zero after the wire is filled to its equilibrium value, simply due to the
lack of a closed electrical circuit. The situation is more interesting in
the case where two leads are coupled to the wire, as it is shown in Fig.\ 
\ref{fig:1}. Fig.\ \ref{fig:2} shows a situation for a wire with two sites,
in which the wire states at time zero are unoccupied. The equilibrium state
is reached after a relaxation time that mostly depends on the wire-lead
coupling strength.  While the number of electrons in the relevant system is
not constant in time, the trace of the reduced density operator is
conserved and normalized. A bias voltage on the system can be simulated by
setting a difference between the left and right Fermi energy, here
$E_{F,l}-E_{F,r}= 10 \Delta$. Since we set the on-site energies to
$E_1=E_2$ and apply a weak coupling scheme where the internal dynamics of
the inner system is about one order faster than the electron transfer
between the leads and the wire, a spatial drop of the population within the
wire in its equilibrium state cannot be observed in the upper panel of
Fig.\ \ref{fig:2}. The equilibrium population of the entire wire is
determined by the average over the expectation values of the left and right
sites, which is given by $\mathrm{tr}(c_1^\dagger c_1
\rho_S(t))=\mathrm{tr}(c_2^\dagger c_2
\rho_S(t))=(n_F(E_1-E_{F,l})+n_F(E_2-E_{F,r}))/2$ if the spectral densities
of both leads, i.e. their coupling to the wire, are the same. The transient
oscillations of the corresponding directed net current decay and a constant
equilibrium value is reached, when the relaxation process is over. The
upper bound for the net current is determined by the small wire-lead
coupling strength, namely the spectral function $J_R(\omega)$ with its
maximum value of $0.1\Delta$.

\subsection{AC voltage}

\begin{figure}
\includegraphics[angle=-90,width=8.5cm,angle=0]{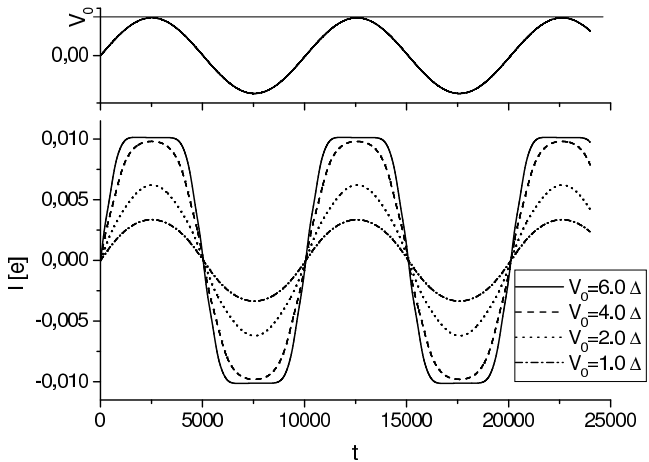}
\caption{The AC voltage (upper panel) with amplitude $V_0$ and frequency $\omega_V=0.00625\Delta$  drives an AC $I[e]$ (bottom panel) between the left and the right lead through  an undriven and not disordered wire consisting of five sites. A current cut-off appears for large AC voltages, since the wire-lead coupling is weak.
\label{fig:3}}
\end{figure}

Time-dependent electronic reservoirs in molecular systems can be modelled
within the Floquet theory by applying a gauge transformation on the wire
Hamiltonian\cite{kohl05a} or by using a source-Redfield\cite{ovch05a}
equation.  Despite the fact that we initially assumed that the lead
Hamiltonian is time-independent, the formalism allows one to treat
time-dependent lead Hamiltonians as well by simply considering
time-dependent coefficients for the correlation functions $C(t)=\sum_k
a_k(t) e^{\gamma_k(t) t}$, in the case of the TL theory, and a time
derivative given by
\begin{eqnarray}
\frac{\mathrm d}{\mathrm dt}C(t)&=& \sum_k^{m+m'}  a_k(t) e^{\gamma_k(t) t}
\left( \gamma_k(t)+ \dot \gamma_k(t) t \right)\nonumber\\
& &+ \sum_k^{m+m'} \dot a_k(t) e^{\gamma_k(t) t}. 
\end{eqnarray}
To get the required form $\dot C(t)=\gamma_k(t)  C(t)$ which is necessary to derive differential equations for Eqs.\ (\ref{equ:aux1}, \ref{equ:aux2}), we have to assume that 
\begin{equation}\label{equ:slow1}
\vert \dot a_k(t)\vert \ll{}\vert a_k(t) \gamma_k(t) \vert,
\end {equation}
\begin{equation} \label{equ:slow2}
\vert e^{\gamma_k(t) t} \dot \gamma_k(t)  t\vert \ll{} \vert e^{\gamma_k(t) t} \gamma_k(t)\vert,
\end{equation}
for all times $t$ not negligible in the time integration of the
auxiliary operators in Eqs.\ (\ref{equ:aux1}, \ref{equ:aux2}). In general,
this assumption corresponds to the statement that the lead dynamics is
sufficiently slower than the lead-wire coupling dynamics. The lead-wire
coupling can be influenced by a variation of the chemical potential of the
electron distribution in the leads, by changing the parameters $\Omega_k,
\Gamma_k, p_k$ of the spectral decomposition, or both. A change in the
chemical potential and keeping the spectral function unchanged would
correspond to a charging process, while shifting the spectral density,
i.e. increasing or decreasing the $\Omega_k$ parameters, and the chemical
potential by the same value describes a total change of the electrostatic
potential without an alteration of the population in the leads. 
In the present work, we restrict ourselves to a time-periodic modulation of the difference between the chemical potentials of the left and right lead $E_{F,r}-E_{F,l}=V_0\, \mathrm{sin} (\omega_V t)$, thereby approximating an AC voltage on the leads with amplitude $V_0$. In this special case, the quality of the approximation of slow dynamics  can be determined considering the general time dependence of the coefficients given by Eqs.\ (\ref{bath12dec}, \ref{bath21dec}) as
$
\vert a_k(t) \vert \sim \vert n_F \left( E_F\left (t \right) \right) \vert
$
for the non-Matsubara terms with the corresponding time-independent exponents $\gamma_k$. Therefore, the time derivative becomes
$
\vert \dot a_k(t) \vert \sim \vert n_F^2 \left( E_F\left (t \right) \right) \beta \dot E_F(t) \vert \sim \vert \frac{V_0 \, \omega_V}{T} \mathrm{cos} \left(\omega_V t \right)  n_F^2 \left( E_F\left (t \right) \right)\vert
$
and conditions (\ref{equ:slow1}, \ref{equ:slow2}) are justified if
$
\vert \frac{V_0 \, \omega_V}{T}\vert \ll{}1
$.
Regarding the Matsubara terms, the relevant time-dependent coefficients are $\gamma_k(t)= -(2 \pi k + \pi) T+ i E_F(t)$ and $\vert a_k(t) \vert = \vert 2 T J(\nu_k)\vert$ for $k=1\, ... \, m'$ and it can be shown that conditions (\ref{equ:slow1}, \ref{equ:slow2}) are also fulfilled if
$
\frac{V_0 \, \omega_V}{ (\pi+2\pi k)  T } \ll{}1
$
which complies with the condition for the non-Matsubara coefficients. A
disregard of the conditions (\ref{equ:slow1}, \ref{equ:slow2}) would cause
additional fast oscillations of the current during those times in which
 the AC voltage changes rapidly.
The AC voltage causes an AC in the system, both shown in Fig.\ \ref{fig:3}, that follows the driving voltage. 
For large voltage amplitudes $V_0$ a finite cut-off appears due to a bottle neck of the system given by its weak coupling to the electronic leads.

\subsection{Coherent destruction of tunneling} \label{subsec:CDT}

\begin{figure}
\includegraphics[angle=-90,width=8.5cm,clip]{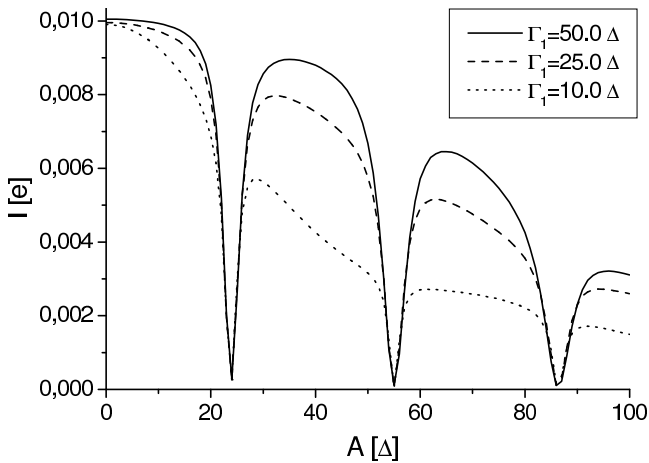}
\caption{The time-averaged current for the periodically driven two-site
  system as a function of the amplitude for different bandwith parameters
  $\Gamma_1$. The frequency of the external field is $\omega_d=10 \Delta$.
  The applied bias voltage is $60 \Delta$. The Fermi energies
  $E_{F,r}=\Omega_1-30 \Delta$, $E_{F,l}=\Omega_1+30 \Delta$ and the
  on-site energies $E_1=E_2=\Omega_1$ are set in relation to the parameter
  $\Omega_1$ of the spectral density.
\label{fig:4}}
\end{figure}

Coherent destruction of tunneling \cite{lehm03a,lehm04a,cama04,gros91,kohl04b} (CDT) is a well understood quantum mechanical effect where an external periodic driving field of the form
\begin{equation}
A(t)=A_0 \, \mathrm{sin} \left(\omega_d t \right)
\end{equation}
yielding an asymmetric potential $U_n(t)=A(t) \delta_{1n}-A(t) \delta_{2n}$ in Eq.\ (\ref{equ:Hamiltonian_dec}) can suppress the time-averaged current in the driven two-site
system depicted in Fig.\ \ref{fig:1}. The current completely breaks down if the ratio of the field parameters  $A_0/\hbar \omega_d$ is equal to a zero of the Bessel function $J_0$ (e.g. 2.405, 5.520, 8.654, ...), a condition that holds for isolated quantum systems as well as for open systems\cite{lehm03a,lehm04a,kohl04b,cama04}.
This is a simple but nontrivial system for a time-dependent conduction
formalism and we extend the results of former calculations which applied the
wide-band
 approximation\cite{lehm03a,lehm04a,kohl04b,cama04} to finite band effects by using a spectral decomposition. The finite width of the decomposed spectral density (\ref{equ:spectralnum}) causes an additional decay of the current with increasing amplitude since the coupling strength between the lead and the corresponding site decreases when the on-site energy of the coupled site is driven to the edges of the Lorentzian during the oscillation generated by the external field $A(t)$. Naturally, this effect becomes more dominant for a smaller bandwidth parameter $\Gamma_1$, shown in Fig.\ \ref{fig:4}. The time-averaged current shows the predicted
breakdowns at the zeros of $J_0$. The amplitude $A_0$ of the applied
electrical field is measured in dimensions of $[\Delta]$ which corresponds
to field strengths of about $10^{8} \frac{\mathrm{V}}{\mathrm{cm}}$ by
assuming intra-atomar distances of $1\mathrm \AA$. The driving frequency is
set to $\omega_d=10 \Delta$ which is in the low energy branch of infrared
light. Here the bias voltage is set to  $E_{F,l}-E_{F,r}$ =  $60 \Delta=6$ eV. 

\begin{figure}
\includegraphics[angle=-90,width=8.5cm,clip]{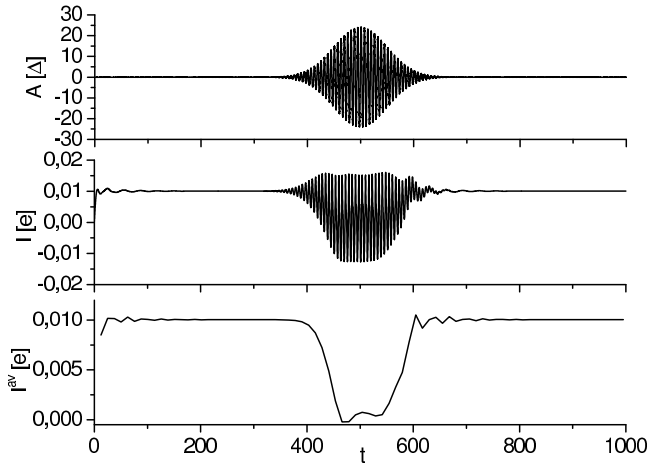}
\caption{The upper panel shows the Gaussian excitation pulse with a peak 
  amplitude of $A_0=24.05 \Delta$ and a width of $\sigma^2=50$. It excites
  the time-resolved net current depicted in the middle panel. If the net
  current is averaged over three periods of the fast pulse oscillation, a
  complete suppression can be observed, depicted in the bottom panel. The
  on-site energies are equal $E_1=E_2=\Omega_1$ and centered between the
  left and the right Fermi energy, which define a DC voltage of
  $E_{F,r}-E_{F,l}= 10 \Delta$.
\label{fig:5}}
\end{figure}

\subsection{Optical current switching} \label{subsec:opt}
The major advantage of the derived conduction formalism, in addition to the avoidance of the wide-band limit, is the applicability to unrestricted time-dependent systems. Former CDT studies of currents in open quantum systems\cite{lehm03a,lehm04a,kohl04b} were based on an infinite-time averaging of the currents due to the mathematical nature of the used approaches. In Fig.\ \ref{fig:5}, we apply a finite laser pulse with a Gaussian shaped amplitude
\begin{equation}
A(t)=A_0 \, \mathrm{exp} \left( \frac{-(t-T)^2}{2 \sigma^2} \right) \mathrm{sin} \left(\omega_d t \right)
\end{equation}
to the asymmetric driven system described in the last subsection. The peak amplitude of the Gaussian was set to $A_0=24.05 \Delta$, where the CDT relation applies. This finite laser pulse causes the time-resolved current to oscillate around zero, shown in the center panel of Fig.\ \ref{fig:5}. The situation becomes more obvious by looking at the time averaged current depicted in the bottom panel of Fig.\ \ref{fig:5}, where the averaging was taken over three oscillations of the driving field. The averaged current is almost suppressed during the time in which the Gaussian laser field is applied to the system and CDT is performed with a finite laser pulse. The effect is superimposed by transient oscillations because the system is under the constraint of permanently changing variables and tries to find its equilibrium state. This finite-time effect gives rise to interesting experimental realizations and further investigations will deal with the idea to apply the current formalism to non-local optimal control algorithms to compute more sophisticated and applicable external control fields.

\section{Summary and conclusion}\label{sec:conclusion}
We have developed a time-dependent non-Markovian conduction formalism based on a projection operator approach
for the density matrix with a numerical decomposition of the spectral density which allows
one to study time-resolved currents and population dynamics in molecular wires for arbitrary time-dependent wire Hamiltonians, which is an useful extension to existing theories and applicable for a wide range of systems.
The formalism includes the coupling to the electronic leads in second-order perturbation theory and a non-perturbative interaction with external fields. Its validity was shown for the example of coherent destruction of tunneling in a driven two-state system and further effects like optical control of current by using a short laser pulse, alternating currents and electrical relaxation of the system into a biased equilibrium state were presented.
These effects were investigated on a femtosecond time scale, which is an important aspect for the the applicability of the formalism for future investigations of femtosecond spectroscopy, optical control and optimization of electrical currents. One can easily include electron-vibrational couplings in second-order perturbation theory and electron-electron correlations resulting in a powerful tool to simulate currents under quite realistic conditions as long as the weak coupling scheme is justified. Furthermore, recent publications\cite{li05,cui05} suggest a way to avoid this weak coupling condition within the framework of the presented formalism.

\acknowledgments 
We are grateful to Markus Schr\"oder and Sigmund Kohler for
useful discussions.


\printfigures

\end{document}